\documentclass[10p]{article}

\usepackage{authblk}
\usepackage{lineno,hyperref}
\usepackage{amsmath}
\usepackage{amssymb}
\usepackage{subcaption}
\usepackage[T1]{fontenc}
\usepackage{graphics,graphicx}
\usepackage{epsfig}
\usepackage{amsbsy}
\usepackage{theorem}
\usepackage{float}
\usepackage{amsfonts}
\usepackage{makeidx}
\usepackage{amsmath}
\usepackage{booktabs}

\author[1,2]{B. Meirbekova}
\author[3]{L. Morini}
\author[4]{M. Brun}
\author[4]{G. Carta}

\affil[1]{Institute of Mathematics and Mathematical Modelling MES RK, Pushkin Street 125, 050010 Almaty, Kazakhstan}
\affil[2]{Department of Computer Engineering, Astana IT University, Mangilik El avenue 55/11, 010000 Astana, Kazakhstan}
\affil[3]{School of Engineering, Cardiff University, The Parade, Cardiff CF24 3AA, UK}
\affil[4]{University of Cagliari, Department of Mechanical, Chemical and Materials Engineering, Cagliari, 09123, Italy}

\title{The strange case of Negative Reflection}

\date{}

\begin{document}

\maketitle


\begin{abstract}
\noindent
In this paper we show for the first time the phenomenon of \emph{negative reflection} in a simple mechanical structure. The latter is a grating of fixed inclusions embedded in a linear elastic matrix. 
Numerical analyses for out-of-plane shear waves demonstrate that there exist frequencies at which most of the incident energy is reflected at negative angles. 
The effect is symmetric with respect to a line that is not parallel to the normal direction to the grating structure. 
Simulations at different angles of incidence and computations of the energy fluxes show that \emph{negative reflection} is achievable in a wide range of loading conditions.
\end{abstract}


\section{Introduction}
An incident wave, impinging on a surface between two continuous media at oblique incidence, generates refracted (or transmitted) and reflected waves. 
The law, attributed to Snell and Descartes \footnote{The study on the historical origins of Snell-Descartes law can be found in \cite{Rashed2000,Kwan2002}. There, the contributions by Ptolemy in the Ancient Greece, by Ibn al-Haytham in the eleventh century and by Kamal al-Din al-Farisi in the fourthteenth century are reported. Two decades before Snell and Descartes, Thomas Harriot and Kepler were already aware of the law of refraction.}, establishes the refraction and reflection conditions\cite{Born1980}. First, the radian frequencies of the transmitted ($\omega_\textup{T}$) and reflected ($\omega_\textup{R}$) waves must be equal to the radian frequency $\omega_\textup{I}$ of the incident monochromatic wave. 
Second, the wave vectors of the transmitted (${\bf k}_\textup{T}$) and reflected (${\bf k}_\textup{R}$) waves must belong to the plane of incidence generated by the incident wave vector ${\bf k}_\textup{I}$ and the normal to the surface ${\bf n}$, and the three wave vectors must have the same component along the surface tangential direction ${\bf t}$ in the plane of incidence, namely ${\bf k}_\textup{I}\cdot{\bf t}={\bf k}_\textup{T}\cdot{\bf t}={\bf k}_\textup{R}\cdot{\bf t}$ (see Figure \ref{Fig01}).  

This leads to the classical form of the Snell-Descartes law: 
\begin{equation}
\label{eqn000}
\frac {\sin \theta _\textup{I}}{c _\textup{I}}=\frac {\sin \theta _\textup{T}}{c _\textup{T}} \,,\qquad  \theta _\textup{I}=\theta _\textup{R}\,,
\end{equation}
where $\theta_\textup{I}$, $\theta_\textup{T}$ and $\theta_\textup{R}$ are the angles of incidence, transmission and reflection measured from the surface normal ${\bf n}$, while $c_\textup{I}$ and $c_\textup{T}$ are the phase velocities of the two media separated by the surface. Accordingly, the rays representing the incident and transmitted waves lie on opposite sides with respect to the line normal to the surface at the point of incidence.
In addition, the direction of the reflected wave is mirrored with respect to the line normal to the surface, namely the angle $\theta_\textup{R}$ is positive.  

By changing the angle of incidence (up to the critical angle) and the phase velocities, it is possible to modulate the amount of transmitted and reflected energy and the phase of the corresponding waves. However, $\theta_\textup{T}$ and $\theta_\textup{R}$ always obey the Snell-Descartes law (\ref{eqn000}).

The equality between the incidence angle $\theta_\textup{I}$ and the reflection angle $\theta_\textup{R}$ does not hold in the vector problem of elasticity or for anisotropic media. In the first case, the tensorial nature of the elastic  problem causes, for any incident wave, the generation of multiple transmitted and reflected waves with longitudinal and transverse polarisations\cite{Achenbach1973} and, for different polarisations of the incident and reflected waves, the angles $\theta_\textup{I}$ and $\theta_\textup{R}$ are different. Furthermore, anisotropies can deviate wave vectors of transmitted and reflected waves\cite{Royer1960,Born1980,Lekner1987}. In both previous cases, the angles of transmission and reflection are constrained by the continuity conditions along the surface.

If the surface between the two media is replaced by a straight homogeneous interface of finite thickness, each single ray is shifted passing through the interface and multiple reflections within the interface generate additional shifted transmitted and reflected waves. Nevertheless, the angles of reflection and transmission continue to obey the laws described above. The same qualitative effect is observed when the interface is a layered structure of homogeneous media\cite{Ewing1953,Brekhovskikh1960}.

We emphasize that, in all the previously considered scenarios, it is not possible to change the signs of the transmission and reflection angles, which remain positive.  

The possibility to obtain negative angles of refraction was postulated in electromagnetism by Veselago\cite{Veselago1968}, who showed that, when both electric permittivity and magnetic permeability are negative, the refractive index is also negative. Such idea was exploited by Pendry\cite{Pendry2000} to propose a first model of perfect lens based on negative refraction. 

The practical implementation of a material with negative refractive index has been realised by developing designs of microstructured interfaces, so that effective negative refraction is attained as a result of a homogenisation process\cite{Smith2000}. Several models leading to negative refraction have been proposed in different fields \cite{Bigoni2013,Colquitt2011,Farhat2008,HladkyHennion2008,Borfiga2019,Morini2019,Chen2022}.

While the problem of negative refraction has been solved, the existence of negative reflection still remains an open problem.
Here, we show that negative angles of reflection can be achieved in a simple mechanical system. We consider propagation of out-of-plane shear waves in a homogenous elastic medium containing an interface made of a repetitive array of fixed holes and we demonstrate numerically the presence of negative reflection.

\section{\label{Sect1}Description of the model}

\subsection{Governing equations}
\label{Sect1.1}

We consider the transmission of time-harmonic out-of-plane shear waves through a grating interface consisting of a square distribution of fixed holes. The displacement vector is given by
\begin{equation}
\label{eqn001}
{\bf u}=
\begin{pmatrix}
0\\
0\\
u({\bf x})
\end{pmatrix}\,,
\end{equation}
where $u({\bf x})$ is the out-of-plane component and ${\bf x}={(x_1\,\,\, x_2)}^{\textup T}$ denotes the in-plane position vector. The dependence on the radian frequency $\omega$ is not indicated for ease of notation. 

For a linear elastic, isotropic and homogeneous medium, the out-of-plane displacement $u$ satisfies the classical Helmholtz equation
\begin{equation}
\label{eqn002}
\Delta u({\bf x}) + \beta^2 u({\bf x}) = 0\,,
\end{equation}
where $\beta=\omega\sqrt{\rho/\mu}$ is the frequency parameter, $\rho$ is the mass density, $\mu$ is the shear modulus and $\Delta$ is the Laplacian operator.

\begin{figure}
\begin{center}
\includegraphics[width=80mm]{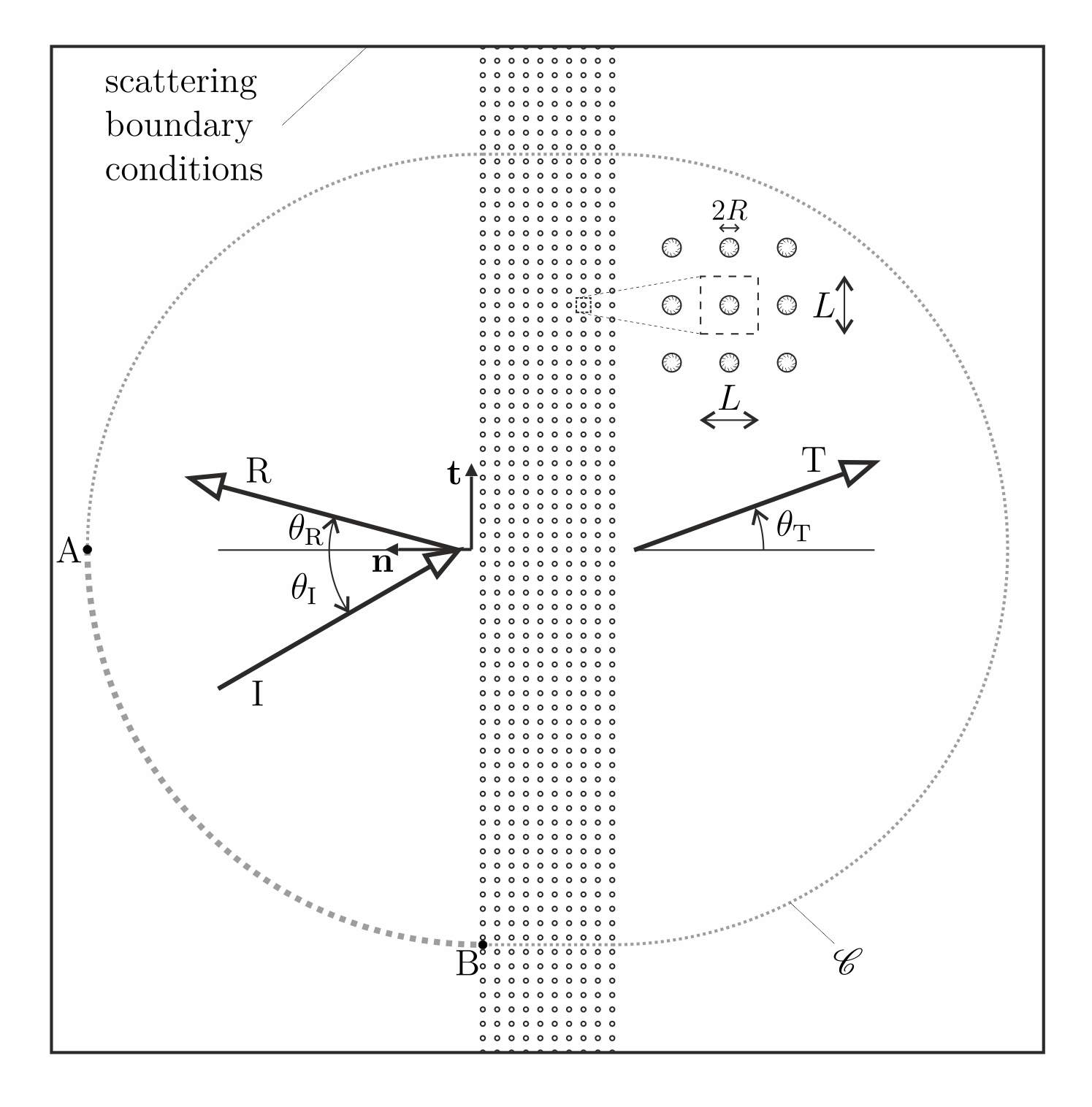}
\end{center}
\caption{\label{Fig01} The mechanical model. The interface consists of a square distribution of fixed circular holes with radius $R=5$ mm. The square cells have side length $L=1$ m. The angles of incidence $\theta_{\textup I}$, reflection $\theta_{\textup R}$ and transmission $\theta_{\textup T}$ of the corresponding waves are indicated. The grey dotted line represents the contour through which energy flux is evaluated.}
\end{figure}

The interface is made of a regular square arrangement of fixed holes, as shown in Figure \ref{Fig01}. Each hole has a circular shape of radius $R=5$ mm; the centres of the holes have positions defined by
\begin{equation}
\label{eqn003}
{\bf  Y}(h_1,h_2)=h_1 {\bf a}^{(1)}+h_2 {\bf a}^{(2)}\,, 
\end {equation}
where ${\bf a}^{(1)}=(L\,\,\,0)^{\textup T}$ and ${\bf a}^{(2)}=(0\,\,\,L)^{\textup T}$ are the lattice vectors, with $L=1$ m. In addition, $(h_1,h_2)^T$ is a multi-index vector\cite{Kittel1956}, with $h_1=0,\,1,\,\ldots,\,9$ and $h_2\in\mathbb Z$.
Dirichlet conditions
\begin{equation}
\label{eqn004}
u(Y_1+R\cos\theta,Y_2+R\sin\theta)=0\,, \qquad 0\le \theta< 2\pi\,,
\end{equation}
are imposed on the boundaries of the holes.

\begin{figure}
\begin{center}
\includegraphics[width=100 mm]{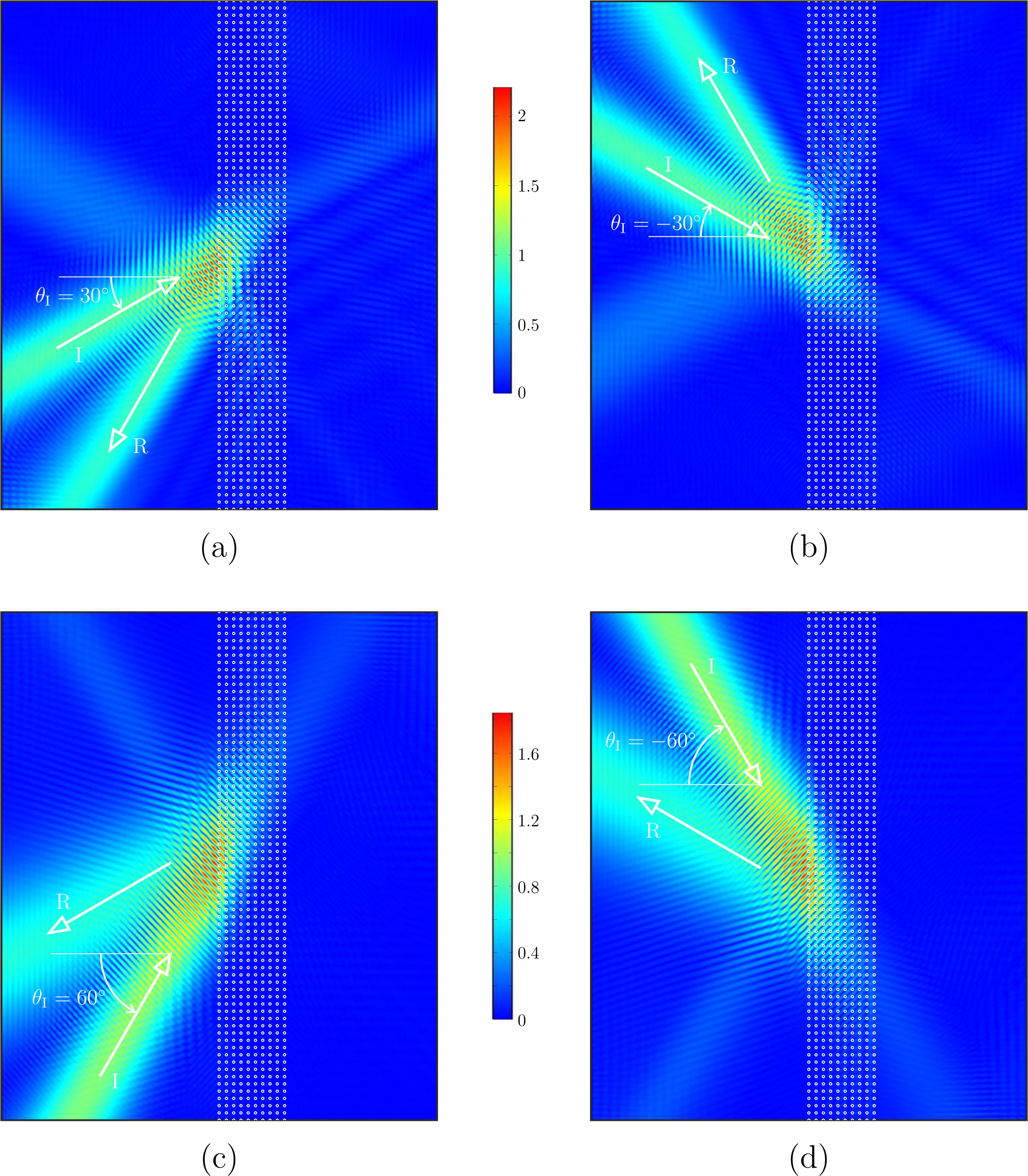}
\end{center}
\caption{\label{Fig02} Negative reflection. Displacement amplitude $|u|$ (in mm) for a Gaussian beam incident at an angle (a) $\theta_\textup{I}=30^\circ$, (b) $\theta_\textup{I}=-30^\circ$, (c) $\theta_\textup{I}=60^\circ$ and (d) $\theta_\textup{I}=-60^\circ$. The shown results are obtained for a frequency parameter $\beta=0.75$ $\mbox{m}^{-1}$.}
\end{figure}

\subsection{Transmission problem}
\label{Sect1.2}

In order to demonstrate negative reflection, we apply an incident Gaussian beam\cite{Svelto2010} that propagates with an incidence angle $\theta_{\textup I}$ and given by
\begin{equation}
\label{eqn005}
u_{\textup I}({\bf x})=U\sqrt{\frac{w_0}{w({\bf m}\cdot{\bf x})}}
\mbox{Exp}\left[\frac{-({\bf p}\cdot{\bf x})^2}{w^2({\bf m}\cdot{\bf x})}-\mathrm{i}\,{\bf k}\cdot{\bf x}-\mathrm{i}\,\beta\frac{({\bf p}\cdot{\bf x})^2}{2R({\bf m}\cdot{\bf x})}+\mathrm{i}\frac{\eta({\bf m}\cdot{\bf x})}{2}  \right]\,.
\end{equation}
In (\ref{eqn005}), ${\bf m}=(\cos\theta_{\textup I}\,\,\,\sin\theta_{\textup I})^{\textup T}$ and ${\bf p}=(-\sin\theta_{\textup I}\,\,\,\cos\theta_{\textup I})^{\textup T}$ are the direction of propagation and its orthogonal one, respectively, while ${\bf k}=\beta {\bf m}$ is the wave vector; furthermore, by fixing the parameters $w_0=5$ m and $x_0=\beta w_0^2/2$, we have:
\begin{eqnarray}
\label{eqn006}
\nonumber
& \displaystyle{w(z) =  w_0 \sqrt{1+\left(\frac{z}{x_0}\right)^2}\,, \qquad 
R(z) = z \left(1 + \frac{x_0^2}{z^2}\right)\,,} \\ & 
\displaystyle{\eta(z)=\arctan\left(\frac{z}{x_0}\right)\,.} 
\end{eqnarray}
Moreover, $U=1$ mm defines the amplitude of the imposed displacement.

The transmission problem is solved by constructing a finite element model in \emph{Comsol Multiphysics} (version 5.6). A finite domain containing $70\times 10$ holes with fixed conditions on their boundaries is implemented. The incident field $u_\textup{I}$ is represented by the Gaussian beam given in (\ref{eqn005}); as a result of the computations, the scattered field $u_\textup{S}$ is determined, so that the total field is $u=u_\textup{I}+u_\textup{S}$.
In the simulations, non-reflecting scattering boundary conditions are applied on the external edges of the finite element rectangular domain (see also Figure \ref{Fig01}).

\begin{figure}
\begin{center}
\includegraphics[width=120 mm]{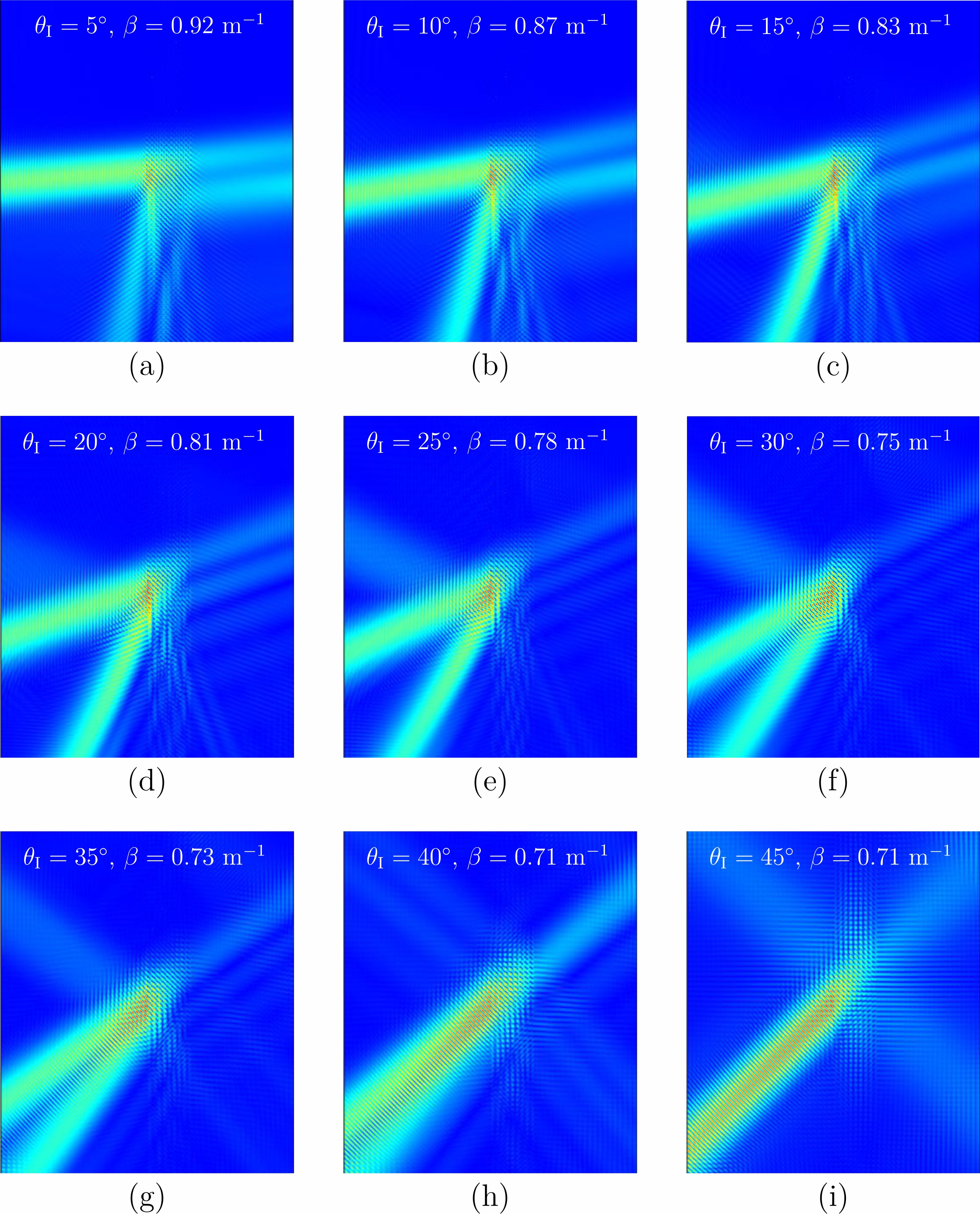}
\end{center}
\caption{\label{Fig03}Amplitudes of the total displacement field for different incidence angles $\theta_\textup{I}$, with $0^\circ \le \theta_\textup{I} \le 45^\circ$, and for different values of the frequency parameter $\beta$.}
\end{figure}

\section{Numerical results}
\subsection{Waveforms exhibiting negative reflection}

The amplitude of the total displacement field $|u|$ is illustrated in Figure \ref{Fig02}. The results are a clear evidence of negative reflection.

By modulating the frequency through the parameter $\beta$, it is possible to detect a reflected and a transmitted beam with $\theta_\textup{R}=\theta_\textup{T}=\theta_\textup{I}$; these waves are predicted by the continuum theory and are visible in Figure \ref{Fig02} as two beams with small amplitude. Additional effects at different frequencies are scattering in multiple directions, shifting of the beams due to the interface, propagation in the interface and negative refraction. These are visible in the video \href{https://clipchamp.com/watch/EHP8HLEM50N}{Negative Reflection}.

In the neighborhood of the two frequencies corresponding to $\beta=0.75$ m$^{-1}$ (Figure \ref{Fig02}) and $\beta=1.47$ m$^{-1}$, a different reflected beam is excited and becomes predominant in terms of scattered energy; this is the one associated with negative reflection.   
In such a case, it is apparent that most of the energy of the incident Gaussian beam is reflected, and the reflected wave is localised within a beam that propagates back with a reflection angle $\theta_\textup{R}$ of opposite sign with respect to that predicted by the Snell-Descartes law.

The effect appears to be reciprocal, so that if the reflection angle is $\theta_\textup{R}=-60^\circ$ when the incidence angle is $\theta_\textup{I}=30^\circ$ (see Figure \ref{Fig02}a), the reflection angle is $\theta_\textup{R}=-30^\circ$ when the incidence angle is $\theta_\textup{I}=60^\circ$ (see Figure \ref{Fig02}c).

Due to the symmetry of the model, the negative reflection phenomenon is retrieved for the negative incidence angles $\theta_\textup{I}=-30^\circ$ (Figure \ref{Fig02}b) and 
$\theta_\textup{I}=-60^\circ$ (Figure \ref{Fig02}d).

In the occurrence of negative reflection, it appears that the bisector line separating incident and reflected waves is not aligned with the direction normal to the interface, but along a direction inclined by $45^\circ$, due to the internal microstructure of the interface. 
This observation is further investigated in Figure \ref{Fig03}, where a set of incident waves with $\theta_\textup{I}=5^\circ,\,10^\circ,\,\ldots,\,45^\circ$ is considered.
For all considered angles $\theta_\textup{I}$, negative reflection occurs with the same bisector line at $45^\circ$. It is also found that maximum reflection is attained at different frequencies: the frequency parameter $\beta$, indicated in Figure \ref{Fig03},
decreases from $\beta=0.92$ m$^{-1}$ for $\theta_\textup{I}=5^\circ$ to $\beta=0.71$ m$^{-1}$ 
for $\theta_\textup{I}=45^\circ$.

\begin{figure}
\begin{center}
\includegraphics[width=75 mm]{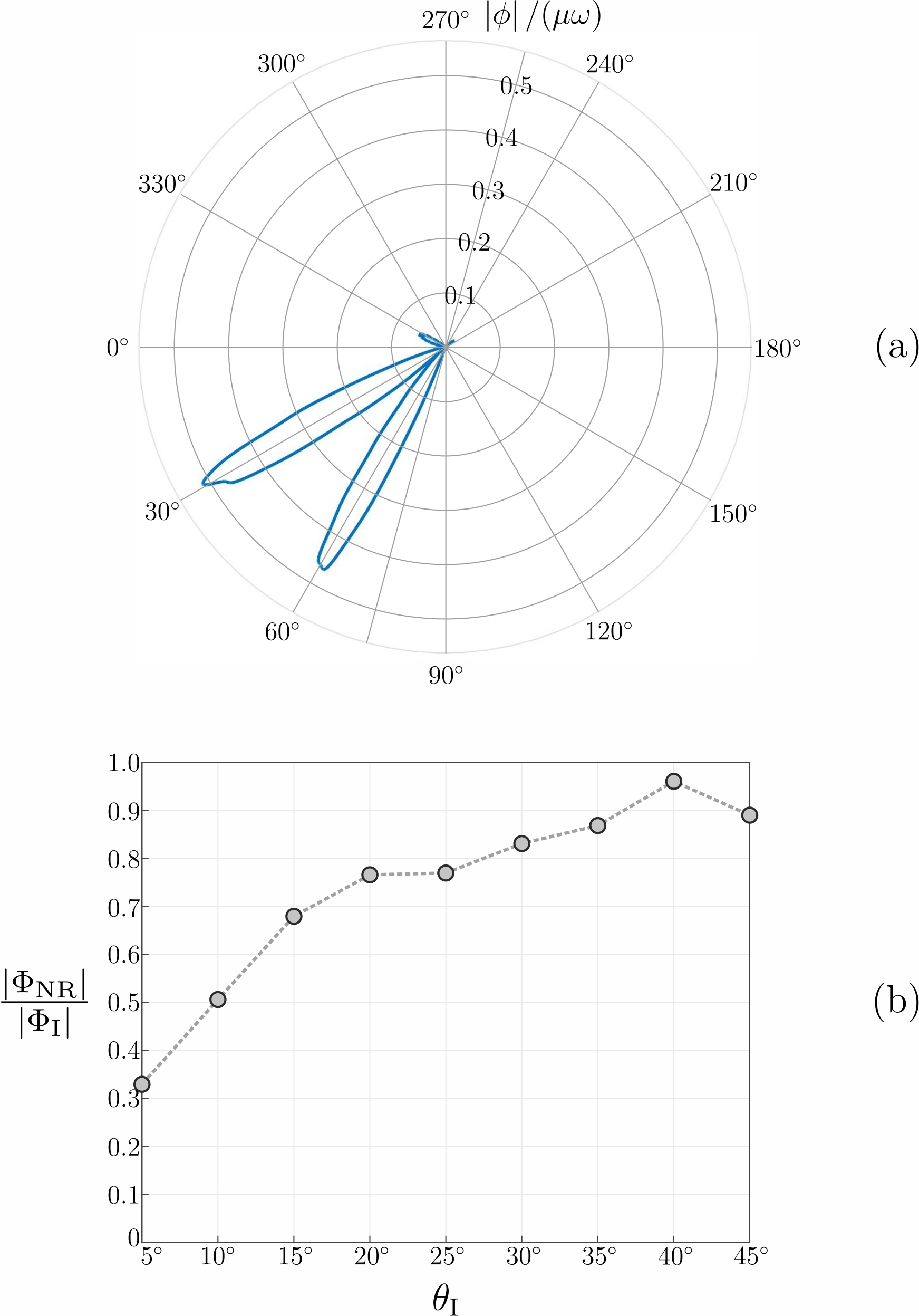}
\end{center}
\caption{\label{Fig04} Energy fluxes. (a) Normalised amplitude $|\phi|/(\mu\omega)$ (in $\mbox{m}^2$) for $\theta_\textup{I}=30^\circ$. (b) Relative total flux amplitude $|\Phi_\textup{NR}|/|\Phi_\textup{I}|$ of the negative reflected beam for different incidence angles $\theta_\textup{I}$.}
\end{figure}

\subsection{Energy flux}

To provide a quantitative measure of the negative reflection phenomenon, we show the diagram of the energy flux on the closed contour $\cal C$ indicated by a dotted grey line in Figure \ref{Fig01}. The contour encompasses 55 rows of holes and it consists of two half circles connected by two straight segments passing through the interface.

The flux, averaged over a period $T=2\pi/\omega$, is \cite{Brillouin1953,Nieves2021}
\begin{equation}
\label{eqn007}
\phi=\frac{1}{2}\mathrm{Re}\left[{\bf \sigma}{\bf n}\cdot{\bf n}\,\, \dot{u}^*\right]=\frac{\mathrm{i}\omega\mu}{2}\mathrm{Im}\left[\frac{\partial u}{\partial n} u^*\right]\,,
\end{equation}
where ${\bf \sigma}$ is the Cauchy stress tensor, ${\bf n}$ the inward normal vector to the curve $\cal C$ and $^*$ denotes the complex conjugate.  

The variation of the flux amplitude $|\phi|$ along the two half circles of the contour $\cal C$, shown in Figure \ref{Fig04}a, evidences that the incident beam at $\theta_\textup{I}=30^\circ$ is reflected and transmitted in three different directions.
It is clear that the energy content of the negative reflected beam is strongly predominant in comparison with the positive reflected and the positive transmitted ones, confirming the occurrence of the negative reflection phenomenon at this frequency.

The ratio between the total flux of the negative reflected beam $|\Phi|_\textup{NR}$ and the total flux of the incident beam $|\Phi|_\textup{I}$ is reported in Figure \ref{Fig04}b.
The flux $\phi$ in Eq. (\ref{eqn007}) is split by introducing the incident and scattered contributions as follows:
\begin{equation}
\label{eqn008}
\begin{split}
\phi=& \frac{\mathrm{i}\omega\mu}{2}\mathrm{Im}\left[\frac{\partial (u_\textup{I}+u_\textup{S})}{\partial n} (u^*_\textup{I}+u^*_\textup{S})\right]\\
= & \frac{\mathrm{i}\omega\mu}{2}\mathrm{Im}\left[\frac{\partial u_\textup{I}}{\partial n} u^*_\textup{I}+
\frac{\partial u_\textup{I}}{\partial n} u^*_\textup{S}+
\frac{\partial u_\textup{S}}{\partial n} u^*_\textup{I}+
\frac{\partial u_\textup{S}}{\partial n} u^*_\textup{S}\right]\\
=&\,\phi_\textup{I}+\phi_\textup{M}+\phi_\textup{S}\,,
\end{split}
\end{equation}
where
\begin{equation}
\label{eqn009}
\phi_\textup{I} = \frac{\mathrm{i}\omega\mu}{2}\mathrm{Im}\left[\frac{\partial u_\textup{I}}{\partial n} u^*_\textup{I}\right]
\end{equation}
is the incident flux,
\begin{equation}
\label{eqn010}
\phi_\textup{S} = \frac{\mathrm{i}\omega\mu}{2}\mathrm{Im}\left[\frac{\partial u_\textup{S}}{\partial n} u^*_\textup{S}\right]
\end{equation}
is the scattered flux and
\begin{equation}
\label{eqn011}
\phi_\textup{M} = \frac{\mathrm{i}\omega\mu}{2}\mathrm{Im}\left[\frac{\partial u_\textup{I}}{\partial n} u^*_\textup{S}+\frac{\partial u_\textup{S}}{\partial n} u^*_\textup{I}\right]
\end{equation}
is the mutual flux.

The total incident flux is obtained as
\begin{equation}
\label{eqn012}
\Phi_\textup{I} = \int_A^B \phi_\textup{I} \, dl \,,
\end{equation}
where the integration is performed between the points $A$ and $B$ along the contour $\cal C$ indicated in Figure \ref{Fig01}.
We identify the total negative reflected flux as 
\begin{equation}
\label{eqn013}
\Phi_\textup{NR} = \int_A^B \phi_\textup{S} \, dl\,.
\end{equation}

The results reported in Figure \ref{Fig04}b indicate that the relative negative reflected energy is above $50\%$ for incidence angles within $35^\circ$ from the bisector line at $45^\circ$, reaching a maximum above $95\%$ for $\theta_\textup{I}=40^\circ$.

In conclusion, we have demonstrated with a series of numerical simulations that negative reflection can occur in a simple mechanical system such as the one considered in this work. The amount of negative reflected energy is relatively large and verified for a wide range of angles of the incident wave. 

\section*{Acknowledgments}
B. Meirbekova acknowledges the financial support of the SC MES RK (Grant No. AP14972863), G. Carta and M.Brun's work has been performed under the auspices of GNFM-INDAM. L. Morini is grateful to the support provided by Cardiff University in the framework of the  scheme  ENGIN  Early  Career Academic Fund 2022.


\bibliographystyle{unsrt}
\bibliography{biblio}

\begin{thebibliography}{10}

\bibitem{Rashed2000}
R.~Rashed.
\newblock A pioneer in anaclastics: {I}bn {S}ahl on burning mirrors and lenses.
\newblock {\em Isis}, 81:464--491, 1990.

\bibitem{Kwan2002}
A.~Kwan, J.~Dudley, and E.~Lantz.
\newblock Who really discovered {S}nell’s law?
\newblock {\em Phys. World}, 15:64, 2002.

\bibitem{Born1980}
M.~Born and E.~Wolf.
\newblock {\em Principles of Optics: Electromagnetic Theory of Propagation,
  Interference and Diffraction of Light}.
\newblock Pergamon Press, Oxford, New York, 1980.

\bibitem{Achenbach1973}
J.~D. Achenbach.
\newblock {\em Wave Propagation in Elastic Solids}.
\newblock McGraw-Hill, New York, 1973.

\bibitem{Royer1960}
D.~Royer and E.~Dieulesaint.
\newblock {\em Elastic Waves in Solids I}.
\newblock Springer, Berlin, 1996.

\bibitem{Lekner1987}
J.~Lekner.
\newblock {\em Theory of Reflection}.
\newblock Martinus Nijhoff Publishers, Dordrecht, 1987.

\bibitem{Ewing1953}
W.~M. Ewing, W.~S. Jardetzky, and F.~Press.
\newblock {\em Elastic Waves in Layered Media}.
\newblock McGraw-Hill Book Company, Inc., New York, Toronto, London, 1957.

\bibitem{Brekhovskikh1960}
L.~Brekhovskikh.
\newblock {\em Waves in Layered Media}.
\newblock Academic Press, Cambridge, Massachusetts, 1960.

\bibitem{Veselago1968}
V.~G. Veselago.
\newblock The electrodynamics of substances with simultaneously negative values
  of $\varepsilon$ and $\mu$.
\newblock {\em Sov. Phys. Usp.}, 10:509, 1968.

\bibitem{Pendry2000}
J.~B. Pendry.
\newblock Negative refraction makes a perfect lens.
\newblock {\em Phys. Rev. Lett.}, 85:3966--3969, 2000.

\bibitem{Smith2000}
D.~R. Smith, W.~J. Padilla, D.~C. Vier, S.~C. Nemat-Nasser, and S.~Schultz.
\newblock Composite medium with simultaneously negative permeability and
  permittivity.
\newblock {\em Phys. Rev. Lett.}, 84:4184--4187, 2000.

\bibitem{Bigoni2013}
D.~Bigoni, S.~Guenneau, A.~B. Movchan, and M.~Brun.
\newblock Elastic metamaterials with inertial locally resonant structures:
  Application to lensing and localization.
\newblock {\em Phys. Rev. B}, 87:174303, 2013.

\bibitem{Colquitt2011}
D.~J. Colquitt, I.~S. Jones, N.~V. Movchan, and A.~B. Movchan.
\newblock Dispersion and localization of elastic waves in materials with
  microstructure.
\newblock {\em Proc. R. Soc. A}, 467:2874--2895, 2011.

\bibitem{Farhat2008}
M.~Farhat, S.~Guenneau, S.~Enoch, G.~Tayeb, N.~V. Movchan, and A.~B. Movchan.
\newblock Analytical and numerical analysis of lensing effect for linear
  surface water waves through a square array of nearly touching rigid square
  cylinders.
\newblock {\em Phys. Rev. E}, 77:046308, 2008.

\bibitem{HladkyHennion2008}
A.~C. Hladky-Hennion, J.~O. Vasseur, G.~Haw, C.~Croënne, L.~Haumesser, and
  A.~N. Norris.
\newblock Negative refraction of acoustic waves using a foam-like metallic
  structure.
\newblock {\em Appl. Phys. Lett.}, 102:144103, 2013.

\bibitem{Borfiga2019}
G.~Bordiga, L.~Cabras, A.~Piccolroaz, and D.~Bigoni.
\newblock Prestress tuning of negative refraction and wave channeling from
  flexural sources.
\newblock {\em Appl. Phys. Lett.}, 114:5084258, 2019.

\bibitem{Morini2019}
L.~Morini, Y.~Eyzat, and M.~Gei.
\newblock Negative refraction in quasicrystalline multilayered metamaterials.
\newblock {\em J. Mech. Phys. Solids}, 124:282--298, 2019.

\bibitem{Chen2022}
Z.~Chen, L.~Morini, and M.~Gei.
\newblock On the adoption of canonical quasi-crystalline laminates to achieve
  pure negative refraction of elastic waves.
\newblock {\em Phil. Trans. R. Soc. A}, 380:20210401, 2022.

\bibitem{Kittel1956}
C.~Kittel.
\newblock {\em Introduction to Solid State Physics}.
\newblock John Wiley and Sons, New York, 1956.

\bibitem{Svelto2010}
O.~Svelto.
\newblock {\em Principles of Lasers}.
\newblock Springer, New York, 2010.

\bibitem{Brillouin1953}
L.~Brillouin.
\newblock {\em Wave Propagation in Periodic Structures. Electric Filters and
  Crystal Lattices}.
\newblock Dover Publication, Inc., New York, 1953.

\bibitem{Nieves2021}
M.~J. Nieves, G.~Carta, V.~Pagneux, and M.~Brun.
\newblock Directional control of {R}ayleigh wave propagation in an elastic
  lattice by gyroscopic effects.
\newblock {\em Front. Mater.}, 7:602960, 2021.

\end{thebibliography}

\end{document}